\documentclass[fleqn,12pt,twoside]{article}
\usepackage{espcrc1,amssymb}

\title{Scalar meson properties from $D$-meson decays\thanks{B.~E. and B.~L. are thankful to the organizers of {\it Few Body 18\/} for 
        their hospitality and support during the conference and acknowledge pleasant and very helpful discussions with J.~P.~B.~C. de Melo, 
        T.~Frederico, D.~Boito, M.~R.~Robilotta and C.~D.~Roberts. Their stay at the Instituto de F\'isica Te\'orica, S\~ao Paulo, before and 
        after the meeting was funded within a joint CNRS--FAPESP agreement. B.~E. is supported by a Marie Curie International 
        Reintegration Grant, No. 516228.}}

\author{B.~El-Bennich\address[LPNHE]{LPNHE (IN2P3--CNRS--Universit\'es Paris 6 et 7), Groupe Th\'eorie \\
        Universit\'e Pierre et Marie Curie,  4 place Jussieu, 75252 Paris, France},
        O.~M.~A.~Leitner\addressmark[LPNHE],
        B.~Loiseau\addressmark[LPNHE]
        and
        J.-P.~Dedonder\address[IMNC]{IMNC, Universit\'e Denis Diderot, case 7021, 2 place Jussieu, 75251 Paris, France}}

\begin{document}
\maketitle

\begin{abstract}
Decay amplitudes of $D(D_s)\to f_0(980)X$, $X=\pi, K$, are compared to experimental branching ratios with the aim of singling 
out the poorly known $D\to f_0(980)$ transition form factor in these amplitudes. Since the other elements of the amplitudes are 
either calculable in an effective QCD theory using operator product expansion or are known from experiment (e.g. the pion and 
kaon decay constants), we can take advantage of these reactions to constrain the transition form factors obtained in relativistic 
quark models~\cite{mesonichp}. In these models, the $f_0(980)$ wavefunction requires an unknown size parameter for both its 
non-strange $\bar uu(\bar dd)$ and strange $\bar ss$ components, which we fit to the $D(D_s)$ decay data.
\end{abstract}

\section{MOTIVATION}

Scalar mesons have been a recurrent topic over the past 30--40 years and while the $\sigma$ has been a longstanding open question since 
the 1960s, the $f_0(980)$ and $a_0(980)$ were firmly established in $\pi\pi$ scattering experiments in the 1970s~\cite{Protopopescu:1973sh}.
The constituent structure of the scalar mesons, nevertheless, has been and still is a controversial issue. 
For a general overview we refer, for instance, to the Particle Data Group review \cite{pdg} and references therein. The known $0^{++}$ mesons
fall into two classes: near and about 1~GeV and in the region $1.3-1.5$~GeV. The scalar objects below 1~GeV form an $SU(3)$ flavor nonet. 
This nonet contains two isosinglets, an isotriplet and two strange isodoublets. Among these lighter scalars, the isosinglet  $f_0(980)$ and the 
isotriplet  $a(980)$ are fairly narrow: $\Gamma=40-100$~MeV. Both couple strongly to the $\bar KK$ channel and lie close to the $\bar KK$ 
threshold at 987~MeV. This closeness to threshold distorts the shape of their resonant structure and the description of the $f_0$ and $a_0$ 
requires a coupled-channel scattering analysis.

The simple quark model views these scalar mesons as orbitally ($L=1$) excited $\bar qq$ states and has been advocated by T\"ornqvist and 
Roos~\cite{Tornqvist:1995ay}. However, some studies \cite{Black:1998wt} tend to favor four-quark configurations of the scalar mesons
and so do coupled-channel analyses~\cite{Krupa:1995fc} or potential models of molecular states strongly coupled to $\pi\pi$ and 
$\bar KK$ thresholds \cite{Weinstein:1990gu}. In any case, the precise four-quark picture is still open to debate since a coupled-channel 
model implies two $\bar qq$ pairs while the configuration $\bar q^2 q^2$ requires two di-quarks.

In the past few years, scalar mesons have emerged in three-body decays of heavy mesons such as $D\to \pi\pi\pi, \pi\pi\bar K, \bar KKK$ 
measured at CESR, DESY and Fermilab~[8--16] and $B\to \pi\pi K, \bar KK K$ decays observed at the dedicated 
$B$-meson facilities BaBar and Belle~\cite{massdistrib}. In particular, in Ref.~\cite{Aitala:2000xu}, an experimental evidence for a light and 
broad scalar resonance in the $m_{\pi\pi}$ spectrum of the $D\to \pi\pi\pi$ decay was found, to be identified with the $f_0(600)$, and also a 
less pronounced $f_0(980)$ peak. Both the BaBar and Belle collaborations~\cite{massdistrib} report a distinctive peak around 1~GeV in the 
$m_{\pi\pi}$ spectrum of three-body $B$-decays.

Our interest in $D$-decays with an $f_0(980)$ in the final state is motivated by two relativistic quark model calculations of $B\to f_0(980)$ and 
$D\to f_0(980)$ transition form factors we have performed recently, one of which makes use of the dispersion relation approach, the other is being
done within Covariant Light-Front Dynamics (CLFD)~\cite{mesonichp}. Both methods are explicitly covariant and require two size parameters 
that specify the unknown $f_0(980)$ wave function. In order to deduce these parameters from experiment, we fit $D(D_s)\to f_0(980)X$ branching 
fractions, where $X$ can be a pion or a kaon. The decay amplitudes are built within the naive QCD factorization approach which neglects 
hard-scattering effects between the spectator and any of the other final quarks. Annihilation topologies are not considered here. Since $m_c$ 
is less than half the value of $m_b$, non-perturbative contributions of order $\Lambda_{\mathrm{QCD}}/m_c$ may be more important than 
in $B$-decay amplitudes and the factorization approach thus less reliable. Nonetheless, already at tree level, non-leptonic two-body 
decay channels of  $D$-mesons can be reasonably reproduced within this simple factorization hypothesis. This is not the case for two-body 
$B$-decays, where penguin contributions are more important. Hence, the decay amplitudes are products of a short-distance part, calculated 
in perturbation theory, and hadronic matrix elements describing the long distance physics such as the pion or kaon decay constant and the 
$D\to f_0(980)$ form factors we are interested in.

\section{DECAY AMPLITUDES IN THE FACTORIZATION APPROACH}

At tree-level order the generic decay amplitudes for a $D$-meson decaying into a scalar $f_0(980)$ and a pseudoscalar meson $P$
are given by
\begin{equation}
  A(D \to f_0 P)=\frac{G_F}{\sqrt{2}}\ a_i f_P (m_D^2-m_{f_0}^2) F_0^{D \to f_0}(m^2_P)\ V_{cq} V_{uq}^*\ .
\end{equation}
Here, $G_F$ is the Fermi coupling constant, $f_P$ is the pseudoscalar decay constant, $V_{cq}$ and $V_{uq}$ are the CKM matrix 
elements with $q=d,s$ and $m_D$ and $m_{f_0}$ are the $D$- and $f_0(980)$-meson masses, respectively. Our main interest here is the 
$D\to f_0(980)$ transition form factor $F_0^{D \to f_0}(m^2_P)$. The short distance physics, beyond the scale $\mu=m_c$, is described 
by the effective coefficients $a_1=C_1+C_2/N_c$ and $a_2=C_2+C_1/N_c$ where $C_1(\mu)$ and $C_2(\mu)$ are the tree-operator 
Wilson coefficients taken at $\mu=m_c$~\cite{Buchalla:1995vs} and $N_c=3$ is the number of colors. The amplitudes for penguin diagrams 
have the same structure, yet their contribution in $D$-decays is considerably smaller and usually neglected.

\begin{table}[t]
\begin{center}
\caption{\small Experimental branching fractions for $D\to f_0(980)X$ and $D_s\to f_0(980)X$ ($X=\pi,K$) deduced from the corresponding
three-body decays [8--16] with $\mathcal{B}(f_0 \to \pi^+ \pi^-) =0.80 \pm 0.14$ and  
$\mathcal{B}(f_0 \to K^+ K^-)=0.11 \pm 0.02$. $\mathcal{F}(D^0 \to f_0 \pi^0)$ is the fit fraction of the $D^0 \to f_0 \pi^0$ decay.}
\label{tab1}
\renewcommand{\tabcolsep}{2pc} 
\renewcommand{\arraystretch}{1.1} 
\begin{tabular}{l|l}\hline 
E791 & $\mathcal{B}(D^+ \to f_0 \pi^+) \times \mathcal{B}(f_0 \to \pi^+ \pi^-)=(1.9 \pm 0.5)\times 10^{-4}$   \\ \hline
FOCUS & $\mathcal{B}(D^+ \to f_0 K^+)  \times\mathcal{B}(f_0 \to K^+ K^-)=(3.84 \pm 0.92)\times 10^{-5}$ \\ 
FOCUS & $\mathcal{B}(D^+ \to f_0 K^+) \times\mathcal{B}(f_0 \to \pi^+ \pi^-)=(6.12 \pm 3.65)\times 10^{-5}$  \\ \hline
ARGUS &  $\mathcal{B}(D^0 \to f_0 \bar{K}^0)\times \mathcal{B}(f_0 \to \pi^+ \pi^-)=(3.2 \pm 0.9)\times 10^{-3}$  \\ 
CLEO &  $\mathcal{B}(D^0 \to f_0 \bar{K}^0) \times\mathcal{B}(f_0 \to \pi^+ \pi^-)=(2.5^{+0.8}_{-0.5})\times 10^{-3}$ \\
BABAR &  $\mathcal{B}(D^0 \to f_0 \bar{K}^0)\times \mathcal{B}(f_0 \to K^+ K^-)=(1.2 \pm 0.9)\times 10^{-3}$  \\ \hline
E687 &   $\mathcal{B}(D_s^+ \to f_0 \pi^+) \times\mathcal{B}(f_0 \to K^+ K^-)=(4.9 \pm 2.3)\times 10^{-3}$ \\
E791 &   $\mathcal{B}(D_s^+ \to f_0 \pi^+)\times \mathcal{B}(f_0 \to \pi^+ \pi^-)=(5.7 \pm 1.7)\times 10^{-3}$ \\
FOCUS &  $\mathcal{B}(D_s^+ \to f_0 \pi^+)\times \mathcal{B}(f_0 \to \pi^+ \pi^-)=(9.5 \pm 2.7)\times 10^{-3}$ \\
FOCUS &  $\mathcal{B}(D_s^+ \to f_0 \pi^+) \times\mathcal{B}(f_0 \to K^+ K^-)=(7.0 \pm 1.9)\times 10^{-3}$  \\ \hline
FOCUS & $\mathcal{B}(D_s^+ \to f_0 K^+) \times\mathcal{B}(f_0 \to K^+ K^-)=(2.8 \pm 1.3)\times 10^{-4}$  \\ \hline
CLEO & $\mathcal{B}(D^0 \to f_0 \pi^0) \simeq  \mathcal{B}(D^0 \to \pi^+ \pi^- \pi^0)\times \mathcal{F}(D^0 \to f_0 \pi^0)=$  \\  
 &  \hspace*{2.6cm} $=(1.1 \pm 0.4)\times10^{-2} \times  (1.0 \pm 0.8)\times10^{-4}$ \\ \hline
\end{tabular}
\end{center}
\vspace*{-0.5cm}
\end{table}

Using the above factorized amplitudes, the decay rates can be computed as
\begin{equation}
 \Gamma(D\to f_0 P) = \frac{1}{8\pi}\frac{|\mathbf{p}|}{m_D^2} \frac{1}{\Gamma_D} |A(D \to f_0 P)|^2\ ,
\label{decayrate}
\end{equation}
where $|\mathbf{p}|= \sqrt{[m_D^{2}-(m_P+m_{f_0})^2][m_D^{2}-(m_P-m_{f_0})^2]}/2m_D$ is the c.m. momentum of the decay particles.
The transition form factor $F_0^{D \to f_0}(m^2_P)$ is related to the $+$ and $-$ components of the pseudoscalar $(P)$ to scalar $(S)$
weak-hadronic matrix element defined as ($p_1$ is the initial and $p_2$ the final momentum and $q=p_1-p_2$)
\begin{equation}
  \langle S(p_2) | J^{\mu}| P(p_1) \rangle = (p_1+p_2)^\mu F_+(q^2) +  (p_1-p_2)^{\mu} F_-(q^2) \ , \quad
  J^\mu = {\bar q} \gamma^\mu (1- \gamma^5)c \ ,
\end{equation}
by the expression
\begin{equation}
F_0(q^2)  =  F_+(q^2) + \frac{q^2}{M^2_1 - M^2_2} F_- (q^2)   \ .
\end{equation}
In our case $M_1$ and $M_2$ are the masses of the $D$- and $f_0(980)$-mesons, respectively. As mentioned, computation of
the $F_+(q^2)$ and $F_-(q^2)$ requires the knowledge of the $f_0(980)$ wavefunction. In Ref.~\cite{mesonichp}, the $f_0(980)$ was
taken to be a $\bar qq$ state with a non-strange and a strange component which implies a mixing angle. Higher Fock states were not 
included. The $f_0(980)$ wavefunction then reads
\begin{equation}
   \Psi_{f_0} = \frac{1}{\sqrt{2}} \Big ( |\bar uu \rangle + |\bar dd \rangle \Big ) \sin \theta_m + |\bar ss \rangle \cos\theta_m = 
   \Phi^n \sin\theta_m +\Phi^s \cos \theta_m,
\label{wavefct}
\end {equation}
where $\theta_m$ is the mixing angle and $n=u(d)$ or $s$. The spatial component $\Phi^n$ and $\Phi^s $ of the wavefunction 
are of Gaussian form $\Phi^{n,s} \propto \exp ( -4\alpha_{n,s} k^2/\mu^2 )$, where $k$ is the modulus of the center-of-mass quark momentum 
and $\mu=m_q m_{\bar q}/(m_q+m_{\bar q})= m_{n,s}/2$  is the reduced mass of the $\bar qq$ pair. The size parameters $\alpha_{n,s}$ are
then fitted via the transition form factor $F_0^{D \to f_0}$ in the decay rates Eq.~(\ref{decayrate}) to the experimental branching ratios listed 
in Table~\ref{tab1}. The mixing angle $\theta_m$ has been estimated from $D_s^+\to f_0(980)\pi^+$ and $D_s^+\to \phi\pi^+$ decays to cover 
the rather wide range $20^\circ \lesssim \theta \lesssim 40^\circ$ and $140^\circ \lesssim \theta \lesssim 160^\circ$.  There are two solutions 
for $\theta_m$ since the mixing angle enters quadratically into the decay rate formula. We choose to determine the mixing angle ourselves 
and thus include $\theta_m$ along with $\alpha_n$ and $\alpha_s$ as a parameter, which we determine by fitting the experimental values in 
Table~\ref{tab1}.

\section{CONCLUSION}

In our effort to model covariant transition form factors for $B\to f_0(980)$ and $D\to f_0(980)$~\cite{mesonichp}, we have used non-leptonic
three-body $D$ decays to fix the parameters $\alpha_n$, $\alpha_s$ and $\theta_m$ of the scalar meson wavefunction. We emphasize that 
our results are strongly dependent on the $\mathcal{B}(f_0 \to \pi^+ \pi^-)$ and  $\mathcal{B}(f_0 \to K^+ K^-)$ fractions and that the uncertainties 
of the experimental branching ratios are large (see Table~\ref{tab1}). Therefore, due to the lack of precision and the need for more data, these 
uncertainties yield large parametric correlations and sizable parameter errors. We hope that newer analyses of three-body $D$-decays 
with higher statistics will remedy this problem. Definite results for $\alpha_{n,s}$ and $\theta_m$ as well as $B\to f_0(980)$ and $D\to f_0(980)$ 
transition form factor predictions for space- and time-like momenta will be communicated promptly.

\end{document}